\documentclass[fleqn,twoside]{article}
\usepackage[headings]{espcrc2}

\readRCS
$Id: espcrc2.tex,v 1.2 2004/02/24 11:22:11 spepping Exp $
\usepackage{graphicx}
\usepackage[figuresright]{rotating}

\newcommand{\AmS}{{\protect\the\textfont2
  A\kern-.1667em\lower.5ex\hbox{M}\kern-.125emS}}

\title{The $\tau$ lepton anomalous magnetic moment}

\author{S. Eidelman\address[NSK]{Budker Institute of Nuclear Physics, 11
  Academician Lavrentiev prospect, Novosibirsk, 630090, Russia}\thanks{Work
  supported in part by the grants of RFBR 06-02-04018 and 06-02-16156 as
  well as by the grant of DFG GZ: 436 RUS 113/769/0-2.},
        M. Giacomini\address[PDU]{Dipartimento di Fisica ``G.~Galilei'', 
          Universit\`{a} di Padova, Via Marzolo 8,
        I-35131, Padova, Italy},
        F.V. Ignatov\addressmark[NSK] 
        and
        M. Passera\addressmark[PDU]\address{INFN, Sezione di Padova, Via
        Marzolo 8, I-35131, Padova, Italy}\thanks{Work supported in part by
        the European Community's Marie Curie Research Training Networks
under contracts
MRTN-CT-2004-503369 and
MRTN-CT-2006-035505. }}

\newcommand{ \be}{\begin{equation}}
\newcommand{ \bm}{\boldmath}
\newcommand{ \ubm}{\unboldmath}
\newcommand{ \ee}{\end{equation}}
\newcommand{ \bea}{\begin{eqnarray}}
\newcommand{ \eea}{\end{eqnarray}}
\newcommand{ \mysmall}[1]{\scriptscriptstyle #1} % a smaller #

\newcommand{ \at}{a_{\tau}}
\newcommand{ \mw}{M_{\mysmall{W}}}
\newcommand{ \mz}{M_{\mysmall{Z}}}
\newcommand{ \mh}{M_{\mysmall{H}}}

\newcommand{ \eq}[1]{Eq.~(\ref{eq:#1})}
\newcommand{ \gev}  {\mbox{ GeV}}
\begin{document}

\begin{abstract}
We review the Standard Model prediction of the $\tau$ lepton $g$$-$$2$
presenting updated QED and electroweak contributions, as well as recent
determinations of the leading-order hadronic term, based on the low energy
$e^+ e^-$ data, and of the hadronic light-by-light one.
\vspace{1pc}
\end{abstract}

\maketitle

%---------------------------------------------------------------------------%
\section{INTRODUCTION}  
%---------------------------------------------------------------------------%
\label{sec:INTRO}

Numerous precision tests of the Standard Model ({\small SM}) and searches
for its possible violation have been performed in the last few decades,
serving as an invaluable tool to test the theory at the quantum level. They
have also provided stringent constraints on many ``New Physics'' ({\small
NP}) scenarios.
A typical example is given by the measurements of the anomalous magnetic
moment of the electron and the muon, $a_e$ and $a_{\mu}$, where recent
experiments reached the fabulous relative precision of 0.7 ppb~\cite{odom}
and 0.5 ppm~\cite{bnl}, respectively (the dimensionless quantity $a$ is
defined as $a = (g - 2)/2$, where $g$ is a gyromagnetic factor).
The anomalous magnetic moment of the electron, $a_e$, is rather insensitive
to strong and weak interactions, hence providing a stringent test of {\small
QED} and leading to the most precise determination of the fine-structure
constant $\alpha$ to date~\cite{Gabrielse_a_2006,MP06}. On the other hand,
the $g$$-$$2$ of the muon, $a_{\mu}$, allows to test the entire {\small SM},
as each of its sectors contributes in a significant way to the total
prediction. Compared with $a_e$, $a_{\mu}$ is also much better suited to
unveil or constrain {\small NP} effects. Indeed, for a lepton $l$, their
contribution to $a_l$ is generally expected to be proportional to
$m_l^2/\Lambda^2$, where $m_l$ is the mass of the lepton and $\Lambda$ is
the scale of {\small NP}, thus leading to an $(m_{\mu}/m_e)^2 \sim 4\times
10^4$ relative enhancement of the sensitivity of the muon versus the
electron anomalous magnetic moment. 
The anomalous magnetic moment of the $\tau$ lepton, $a_{\tau}$, would suit even
better; however, its relatively short lifetime makes a direct measurement
impossible, at least at present.

Recent high-precision experiments at low-energy $e^+e^-$
colliders~\cite{cmd2rho,sndff,kloe} allowed a significant improvement of the
uncertainty of the leading-order hadronic contribution to
$a_{\mu}$~\cite{se06,kaoru06}.  In parallel to these efforts, many other
improvements of the {\small SM} prediction for $a_{\mu}$ were carried on in
recent years (see Refs.~\cite{g-2_mureviews} for reviews). All these
experimental and theoretical developments allow to significantly improve the
{\small SM} prediction for the anomalous magnetic moment of the $\tau$
lepton as well, which is usually split into three parts: {\small QED},
electroweak and hadronic (see~\cite{pas06,EP06} for a very recent review).

%---------------------------------------------------------------------------%
\section{QED CONTRIBUTION TO \bm $\at$ \ubm}
%---------------------------------------------------------------------------%
\label{sec:QED}

The {\small QED} contribution to $a_{\tau}$ arises from the subset of
{\small SM} diagrams containing only leptons ($e,\mu,\tau$) and photons. The
leading (one-loop) contribution was first computed by Schwinger more than
fifty years ago~\cite{Sch48}.  Also the two- and three-loop {\small QED}
terms are known (see Refs.~\cite{MP06,EP06}).
%while higher-order contributions are not, contrary to the case of the
%electron and muon $g$$-$$2$~\cite{QED_4E,QED45}. (An exception is the mass-
%and flavor-independent four-loop term $A_4^{(8)}$~\cite{QED_4E}, which is
%however expected to be a very small part of the complete four-loop
%contribution.)  
Adding up these contributions and using the latest {\small
CODATA}~\cite{CODATA02} recommended mass ratios $m_{\tau}/m_e = 3477.48
(57)$ and $m_{\tau}/m_{\mu} = 16.8183 (27)$, and new value of $\alpha$
derived in Refs.~\cite{Gabrielse_a_2006} and \cite{MP06},
$\alpha^{-1} \, = \,  137.035 \, 999 \, 709 \, (96)$,
one obtains the total {\small QED} contribution to $a_{\tau}$~\cite{MP06}:
\be
    a_{\tau}^{\mysmall \rm QED} =
    117 \, 324 \, (2) \times 10^{-8}.
\label{eq:TQED}
\ee
The error $\delta a_{\tau}^{\mysmall \rm QED}$ is the uncertainty
$\delta C_{\tau}^{(8)}(\alpha/\pi)^4 \sim \pi^2 \ln^2(m_{\tau}/m_e)
(\alpha/\pi)^4 \sim 2\times 10^{-8}$
assigned in \cite{MP06} to $a_{\tau}^{\mysmall \rm QED}$ for uncalculated
four-loop contributions. The errors due to the uncertainties of the
$O(\alpha^2)$ ($5 \times 10^{-10}$) and $O(\alpha^3)$ terms ($3 \times
10^{-11}$), as well as that induced by the uncertainty of $\alpha$ ($8
\times 10^{-13}$) are negligible.
The result in \eq{TQED} supersedes the earlier values $a_{\tau}^{\mysmall
\rm QED} = 117 \, 319 \, (1) \times 10^{-8}$~\cite{Samuel_tau} and
$a_{\tau}^{\mysmall \rm QED} = 117 \, 327.1 \, (1.2) \times
10^{-8}$~\cite{Narison01} (see Refs.~\cite{MP06,EP06} for details).

%---------------------------------------------------------------------------%
\section{EW CONTRIBUTION TO \bm $\at$ \ubm}
%---------------------------------------------------------------------------%
\label{sec:EW}

With respect to Schwinger's contribution, the electroweak ({\small EW})
correction to the anomalous magnetic moment of the $\tau$ lepton is
suppressed by the ratio $(m_{\tau}/\mw)^2$, where $\mw$ is the mass of the
$W$ boson. Numerically, this contribution is of the same order of magnitude
as the three-loop {\small QED} one.

%---------------------------------------------------------------------------%
\subsection{One-loop Contribution}
\label{subsec:EW1}

The one-loop {\small EW} term is~\cite{ew1loop}:
$
     a_{\tau}^{\mysmall \rm EW} (\mbox{1 loop}) = 
     \frac{5 G_{\mu} m^2_{\tau}}{24 \sqrt{2} \pi^2}
     \left[ 1+ \frac{1}{5}\left(1-4\sin^2\!\theta_{\mysmall{W}}\right)^2 
       + O(m^2_{\tau}/M^2_{\mysmall{Z,W,H}}) \right],
$
where $G_{\mu}=1.16637(1) \times 10^{-5}\gev^{-2}$ is the Fermi coupling
constant~\cite{PDG06}, $\mz$, $\mw$ and $\mh$ are the masses of the $Z$, $W$
and Higgs bosons, and $\theta_{\mysmall{W}}$ is the weak mixing angle.
Closed analytic expressions for $\at^{\mysmall \rm EW} (\mbox{1 loop})$
taking exactly into account the $m^2_{\mu}/M^2_{\mysmall{B}}$ dependence
($B=Z,W,$ Higgs, or other hypothetical bosons) can be found in
Ref.~\cite{Studenikin}.  Employing the on-shell definition
$\sin^2\!\theta_{\mysmall{W}} =
1-M^2_{\mysmall{W}}/M^2_{\mysmall{Z}}$~\cite{Si80}, where
$\mz=91.1875(21)\gev$ and $\mw$ is the {\small SM} prediction of the $W$
mass (which can be derived, for example, from the simple formulae of
\cite{Formulette} leading to $\mw =80.383\gev$ for $\mh=150\gev$), and
including the tiny $O(m^2_{\tau}/M^2_{\mysmall{Z,W,H}})$ corrections of
Ref.~\cite{Studenikin}, for $\mh=150\gev$ one obtains
$ \at^{\mysmall \rm EW} (\mbox{1 loop}) = 55.1(1) \times
    10^{-8}~\cite{EP06}.
\label{eq:EWoneloopN}
$
The uncertainty encompasses the shifts induced by variations of $\mh$ from
114.4 GeV, the current lower bound at 95\% confidence level~\cite{LEPHIGGS},
up to a few hundred GeV, and the tiny uncertainty due to the error in
$m_{\tau}$.
The estimate of the total {\small EW} contribution of
Ref.~\cite{Samuel_tau}, $\at^{\mysmall \rm EW} = 55.60(2) \times 10^{-8}$,
obtained from the one-loop formula (without the small corrections of order
$m^2_{\tau}/M^2_{\mysmall{Z,W,H}}$), is similar to the one-loop value
reported above. However, its uncertainty ($2 \times 10^{-10}$) is too small,
and it doesn't contain the two-loop contribution which, as we'll now
discuss, is not negligible.

%---------------------------------------------------------------------------%
\subsection{Two-loop Contribution}
\label{subsec:EW2}

The two-loop {\small EW} contributions $a_l^{\mysmall \rm EW} (\mbox{2
loop})$ ($l\!=\!e$, $\mu$ or $\tau$) were computed in 1995 by Czarnecki,
Krause and Marciano~\cite{CKM95a,CKM95b,CK96}.  This remarkable calculation
leads to a significant reduction of the one-loop prediction because of large
factors of $\ln(M_{\mysmall{Z,W}}/m_f)$, where $m_f$ is a fermion mass scale
much smaller than $\mw$~\cite{KKSS}.

The two-loop contribution to $\at^{\mysmall \rm EW}$ can be divided into
fermionic and bosonic parts; the former, $\at^{\mysmall \rm EW}(\mbox{2 loop
fer})$, includes all two-loop {\small EW} corrections containing closed
fermion loops, whereas all other contributions are grouped into the latter,
$\at^{\mysmall \rm EW}(\mbox{2 loop bos})$. The bosonic part was computed in
Ref.~\cite{CKM95b}; its value, for $\mh=150\gev$, is
$\at^{\mysmall \rm EW}(\mbox{2 loop bos}) = -3.06 \times
10^{-8}$~\cite{EP06}.

The fermionic part of $\at^{\mysmall \rm EW}(\mbox{2 loop})$ also contains
the contribution of diagrams with light quarks; they involve long-distance
{\small QCD} for which perturbation theory cannot be employed. In
particular, these hadronic uncertainties arise from two types of two-loop
diagrams: those with the hadronic photon--$Z$ mixing, and those containing
quark triangle loops with the external photon, a virtual photon and a $Z$
attached to them.  The hadronic uncertainties mainly arise from the latter
ones.  In Refs.~\cite{CKM95a,CK96} these nonperturbative effects were
modeled introducing effective quark masses as a simple way to account for
strong interactions, and the diagrams with quark triangle loops were
computed with simple approximate expressions.  However, contrary to the case
of the muon $g$$-$$2$, where all fermion masses (with the exception of
$m_e$) enter in the evaluation of these contributions, in the case of the
$\tau$ lepton these approximate expressions do not depend on any fermion
mass lighter than $m_{\tau}$; apart from $m_{\tau}$, they only depend on
$M_{\rm\scriptstyle top}$ and $m_{b}$, the masses of the top and bottom
quarks (assuming the charm mass $m_c\!<\!m_{\tau}$).  Very recently, this
analysis was slightly refined in Ref.~\cite{EP06} by numerically integrating
exact expressions provided in Ref.~\cite{KKS} for arbitrary values of $m_f$,
obtaining, for $\mh\!=\!150\gev$,
$\at^{\mysmall \rm EW}(\mbox{2 loop fer}) = -4.68 \times
10^{-8}$~\cite{EP06}.
This evaluation also included the tiny $O(10^{-9})$ contribution of the
$\gamma$--$Z$ mixing diagrams, suppressed by
($1-4\sin^2\!\theta_{\mysmall{W}}) \!\sim\! 0.1$ for quarks and
($1-4\sin^2\!\theta_{\mysmall{W}})^2$ for leptons, via the explicit formulae
of Ref.~\cite{CMV03}.

The sum of the fermionic and bosonic two-loop {\small EW} contributions
described above gives
$\at^{\mysmall \rm EW}(\mbox{2 loop}) \!=\! -7.74 \times
10^{-8}$~\cite{EP06},
a 14\% reduction of the one-loop result.  The leading-logarithm three-loop
{\small EW} contributions to the muon $g$$-$$2$ were determined to be
extremely small via renormalization-group analyses~\cite{CMV03,DGi98}. In
Ref.~\cite{EP06} an additional uncertainty of
$O[\at^{\mysmall \rm EW}(\mbox{2 loop}) (\alpha/\pi)
\ln(\mz^2/m_{\tau}^2)] \!\sim\! O(10^{-9})$
was assigned to $\at^{\mysmall \rm EW}$ to account for these neglected
three-loop effects. Adding $\at^{\mysmall \rm EW}(\mbox{2 loop})$ to the
one-loop value presented above, one gets the total {\small
EW} correction (for $\mh\!=\!150\gev$)~\cite{EP06}:
\be
    \at^{\mysmall \rm EW} = 47.4 (5) \times 10^{-8}.
\label{eq:TEW}
\ee
The uncertainty allows $\mh$ to range from 114 GeV up to $\sim \! 300$ GeV,
and reflects the estimated errors induced by hadronic loop effects,
neglected two-loop bosonic terms and the missing three-loop contribution. It
also includes the tiny errors due to the uncertainties in $M_{\rm\scriptstyle
top}$ and $m_{\tau}$. The value in \eq{TEW} is in agreement with the
prediction
  $\at^{\mysmall \rm EW} = 47 (1) \times 10^{-8}$~\cite{CK96,Narison01},
with a reduced uncertainty.  As we mentioned in Sec.~\ref{subsec:EW1}, the
{\small EW} estimate of Ref.~\cite{Samuel_tau},
$\at^{\mysmall \rm EW} = 55.60(2) \times 10^{-8}$, mainly differs from
\eq{TEW} in that it doesn't include the two-loop corrections.

%---------------------------------------------------------------------------%
\section{THE HADRONIC CONTRIBUTION}
%---------------------------------------------------------------------------%
\label{sec:HAD}

In this section we will analyze $a_{\tau}^{\mysmall \rm HAD}$, the
contribution to the $\tau$ anomalous magnetic moment arising from {\small
QED} diagrams involving hadrons. Hadronic effects in (two-loop) {\small EW}
contributions are already included in $\at^{\mysmall \rm EW}$ (see
Sec.~\ref{sec:EW}).

%---------------------------------------------------------------------------%
\subsection{Leading-order Hadronic Contribution}
\label{subsec:HLO}

Similarly to the case of the muon $g$$-$$2$, the leading-order hadronic
contribution to the $\tau$ lepton anomalous magnetic moment is given by the
dispersion integral~\cite{DispInt}:
\begin{equation}
{a}^{\mysmall \rm HLO}_{\tau} \!=\!
 \frac{m^2_\tau}{12\pi^3}
\!\int_{4m^2_\pi}^{\infty} \! \! \! ds\: 
\frac{\sigma^{(0)}\!(e^+e^- \!\! \to \! {\rm hadrons}) K_{\tau}(s)}{s},
\label{eq:dispint}
\end{equation}
in which the role of the low energies is very important, although not as
strongly as in $a_{\mu}^{\mysmall \rm HLO}$.

The history of the $a_{\tau}$ 
calculations~\cite{Samuel_tau,Narison01,nar78,bs88,ej95,j96,EP06} based 
mainly on experimental $e^+e^-$ data is shown in Table~\ref{tab:atau}.   
Purely theoretical estimates somewhat undervalue the hadronic
contribution and have rather large 
uncertainties~\cite{ben93,hn93,hold,dor}.
%The history of the $a_{\tau}$ calculations, which started in 
%1978~\cite{nar78}, is shown in Table~\ref{tab:atau}. Its first part 
%contains those based mainly on experimental $e^+e^-$ data,  
%while its second part shows
%purely theoretical estimates~\cite{ben93,hn93,hold,dor},
%which somewhat undervalue the hadronic
%contribution and have rather large uncertainties.
% is not as rich as that of the muon and  
%The first calculation performed in 1978 in Ref.~\cite{nar78} was based on
%experimental data available at that time below 7.4~GeV, whereas at higher
%energies the asymptotic {\small QCD} prediction was used.  Ten years
%later, a rough estimate was made in Ref.~\cite{bs88} based on low energy
%$e^+e^-$ data.  In Ref.~\cite{Samuel_tau} the contribution of the $\rho$
%meson was estimated by integrating the approximation obtained using the
%Breit-Wigner curve, while other contributions used the data. The accuracy of
%the calculation was considerably improved in Refs.~\cite{ej95,j96} where,
%below 40~GeV, only data were used. In Ref.~\cite{Narison01}, data were
%only used below 3~GeV (together with the experimental parameters of the
%$J/\psi$ and $\Upsilon$ family states). In our opinion this can
%significantly underestimate the resulting uncertainty. In addition, in the
%same reference, data from $\tau$ lepton decays were extensively used; as it
%is known today, this leads to higher spectral functions than in $e^+e^-$
%case,\cite{dehz1,dehz2} and can therefore overestimate the result. The
%results of these calculations are summarized in Table~\ref{tab:atau}.
%% 

% 
% TABLE
\vspace*{-5mm}

\begin{table}[htb]
\caption{Calculations of ${\rm a}_{\tau}^{\mysmall \rm HLO}$}
\begin{tabular}{@{}lc@{}} 
\hline
Author & ${\rm a}_{\tau}^{\mysmall \rm HLO} \times 10^{8}$  \\
\hline
Narison \cite{nar78} & 370 $\pm$ 40 \\
Barish \& Stroynowski~\cite{bs88} & $\sim 350$ \\
Samuel et al. \cite{Samuel_tau}  &  $360 \pm 32$ \\
Eidelman \& Jegerlehner \cite{ej95,j96}  & 
$338.4 \pm 2.0 \pm 9.1$ \\
Narison \cite{Narison01} &  353.6 $\pm$ 4.0 \\
Eidelman \& Passera~\cite{EP06}   & 337.5 $\pm$ 3.7 \\
%\hline
%Benmerrouche et al.~\cite{ben93}  & 197--246          \\
%Hamzeh \& Nasrallah \cite{hn93}  & $280 \pm 20$ \\
%Holdom et al.~\cite{hold}  & $320 \pm 10$ \\
%Dorokhov~\cite{dor} & $310 \pm 20$ \\
\hline
\end{tabular}
\label{tab:atau}
\end{table}
%
%For completeness, its second part also shows
%purely theoretical estimates~\cite{ben93,hn93,hold,dor},
% The analysis based on {\small QCD} sum rules
%performed in Ref.~\cite{ben93} gives results strongly depending on the
%choice of quark and gluon condensates. {\small QCD} sum rules are also used
%in Ref.~\cite{hn93}.  In Ref.~\cite{hold} the authors use a nonlocal
%constituent quark model for the description of the photon vacuum
%polarization function $\Pi_{\rm had}(q^2)$ at space-like momenta and obtain
%$a_{\tau}^{\mysmall \rm HLO} = 3.2(1) \times 10^{-6}$, close to the
%estimates based on the experimental data. They also show that a simpler
%model with constituent quark masses independent of momentum is strongly
%dependent on the values chosen for the quark masses. For example, with
%$m_u=m_d=330$~MeV and $m_s=$~550~MeV their result is $a_{\tau}^{\mysmall \rm
%HLO} = 2.2(1) \times 10^{-6}$, i.e., significantly smaller than the previous
%estimate. They could reproduce the value $a_{\tau}^{\mysmall \rm HLO} =
%3.2(1) \times 10^{-6}$ using $m_u=m_d=m_s=201$~MeV.  In a recent analysis
%using the instanton liquid model the author obtains $(3.1 \pm 0.2) \times
%10^{-6}$.~\cite{dor} 
%which somewhat undervalue the hadronic
%contribution and have rather large uncertainties.
\vspace*{-3mm}

We updated the calculation of the leading-order contribution using the whole
bulk of experimental data below 12~GeV, which include old data compiled in
Refs.~\cite{ej95,dehz}, recent results from the {\small CMD-2} and {\small
SND} detectors in Novosibirsk~\cite{cmd2rho,sndff,se06}, and from the 
radiative return studies at {\small KLOE} in Frascati~\cite{kloe} and 
BaBar at {\small SLAC}~\cite{babr}. The improvement is particularly strong 
in the channel
$e^+e^- \to\pi^+\pi^-$.
Our result is
\be
    a_{\tau}^{\mysmall \rm HLO} =  337.5 \, (3.7) \times 10^{-8}
\label{eq:THLO}.
\ee
The overall uncertainty is 2.5 times smaller than
that of the previous data-based prediction~\cite{ej95,j96}.    

%---------------------------------------------------------------------------%
\subsection{Higher-order Hadronic Contributions}
\label{subsec:HHO}

The hadronic higher-order $(\alpha^3)$ contribution $a_{\tau}^{\mysmall \rm
HHO}$ can be divided into two parts:
$
     a_{\tau}^{\mysmall \rm HHO}=
     a_{\tau}^{\mysmall \rm HHO}(\mbox{vp})+
     a_{\tau}^{\mysmall \rm HHO}(\mbox{lbl}).
$
The first one is the $O(\alpha^3)$ contribution of diagrams containing
hadronic self-energy insertions in the photon propagators.  It was
determined by Krause in 1996~\cite{Krause96}: 
\be
a_{\tau}^{\mysmall \rm HHO}(\mbox{vp})= 7.6 (2) \times 10^{-8}.
\label{eq:THHOVAC}
\ee
Note that na\"{\i}vely rescaling the muon result by the factor
$m_{\tau}^2/m_{\mu}^2$ (as it was done in Ref.~\cite{Samuel_tau}) leads
to the totally incorrect estimate $a_{\tau}^{\mysmall \rm HHO}(\mbox{vp})=
(-101\times 10^{-11}) \times m_{\tau}^2/m_{\mu}^2 = -29 \times 10^{-8}$ (the
$a_{\mu}^{\mysmall \rm HHO}(\mbox{vp})$ value is from
Ref.~\cite{Krause96}); even the sign is wrong!

The second term, also of $O(\alpha^3)$, is the hadronic light-by-light
contribution. Similarly to the case of the muon $g$$-$$2$, this term cannot
be directly determined via a dispersion relation approach using data (unlike
the leading-order hadronic contribution), and its evaluation therefore
relies on specific models of low-energy hadronic interactions with
electromagnetic currents. Until recently, very few estimates of
$a_{\tau}^{\mbox{$\scriptscriptstyle{\rm HHO}$}}(\mbox{lbl})$ existed in the
literature~\cite{Samuel_tau,Krause96,Narison01}, and all of them were
obtained simply rescaling the muon results $a_{\mu}^{\mysmall \rm
HHO}(\mbox{lbl})$ by a factor $m_{\tau}^2/m_{\mu}^2$.
Following this very na\"{\i}ve procedure, the $a_{\tau}^{\mysmall \rm
HHO}(\mbox{lbl})$ estimate varies between
$     a_{\tau}^{\mysmall \rm HHO}(\mbox{lbl}) = 
%[80(40) \times 10^{-11}]\times (m_{\tau}^2/m_{\mu}^2)= 
23(11)\times 10^{-8}
$, 
and
$     a_{\tau}^{\mysmall \rm HHO}(\mbox{lbl}) = 
%[136(25) \times 10^{-11}]\times (m_{\tau}^2/m_{\mu}^2)= 
38(7)\times 10^{-8}
$, 
according to the values chosen for $a_{\mu}^{\mysmall \rm HHO}(\mbox{lbl})$
from Refs.~\cite{lbl} and \cite{MV03}, respectively.

These very na\"{\i}ve estimates fall short of what is needed. Consider the
function $A_2^{(6)}(m_l/m_j,\mbox{lbl})$, the three-loop {\small QED}
contribution to the $g$$-$$2$ of a lepton of mass $m_l$ due to
light-by-light diagrams involving loops of a fermion of mass $m_j$. 
This function was computed in Ref.~\cite{LR93} for
arbitrary values of the mass ratio $m_l/m_j$. In particular, if $m_j \gg
m_l$, then $A_2^{(6)}(m_l/m_j,\mbox{lbl}) \sim (m_l/m_j)^2$. This implies
that, for example, the (negligible) part of $a_{\tau}^{\mysmall \rm
HHO}(\mbox{lbl})$ due to diagrams with a top-quark loop can be reasonably
estimated simply rescaling the corresponding part of $a_{\mu}^{\mysmall \rm
HHO}(\mbox{lbl})$ by a factor $m_{\tau}^2/m_{\mu}^2$. On the other hand, to
compute the dominant contributions to $a_{\tau}^{\mysmall \rm
HHO}(\mbox{lbl})$, i.e.\ those induced by the light quarks, we need the
opposite case: $m_j \ll m_l = m_{\tau}$. In this limit,
$A_2^{(6)}(m_l/m_j,\mbox{lbl})$ does not scale as $(m_l/m_j)^2$, and a
na\"{\i}ve rescaling of $a_{\mu}^{\mysmall \rm HHO}(\mbox{lbl})$ by
$m_{\tau}^2/m_{\mu}^2$ to derive $a_{\tau}^{\mysmall \rm HHO}(\mbox{lbl})$
leads to an incorrect estimate.

For these reasons, a parton-level estimate of $a_{\tau}^{\mysmall \rm
HHO}(\mbox{lbl})$ was recently performed in Ref.~\cite{EP06}, based on the
exact expression for $A_2^{(6)}(m_l/m_j,\mbox{lbl})$, using the quark masses
recently proposed in Ref.~\cite{ES06} for the determination of
$a_{\mu}^{\mysmall \rm HHO}(\mbox{lbl})$: $m_u=m_d=176$ MeV, $m_s=305$ MeV,
$m_c=1.18$ GeV and $m_b=4$ GeV (note that with these values the authors of
Ref.~\cite{ES06} obtain $a_{\mu}^{\mysmall \rm HHO}(\mbox{lbl}) = 136 \times
10^{-11}$, in perfect agreement with the value in Ref.~\cite{MV03} -- see
also Ref.~\cite{Pivovarov03} for a similar earlier determination). The
result of this analysis is~\cite{EP06}
\be
a_{\tau}^{\mysmall \rm HHO}(\mbox{lbl})= 5 (3) \times 10^{-8}.
\label{eq:THHOLBL}
\ee
This value is much lower than those obtained by simple rescaling above. 
%of $a_{\mu}^{\mysmall \rm HHO}(\mbox{lbl})$ by $m_{\tau}^2/m_{\mu}^2$.  
The dominant contribution comes from the $u$ quark; the uncertainty $\delta
a_{\tau}^{\mysmall \rm HHO}(\mbox{lbl})= 3 \times 10^{-8}$ allows $m_u$ to
range from 70 MeV up to 400 MeV. Further independent studies (following the
approach of Ref.~\cite{MV03}, for example) would provide an important
check of this result.

The total hadronic contribution to the anomalous magnetic moment of the
$\tau$ lepton can be immediately derived adding the values in
Eqs.~(\ref{eq:THLO}), (\ref{eq:THHOVAC}) and (\ref{eq:THHOLBL}),
\bea
      a_{\tau}^{\mysmall \rm HAD} &=& a_{\tau}^{\mysmall \rm HLO}+
      a_{\tau}^{\mysmall \rm HHO}(\mbox{vp})+
      a_{\tau}^{\mysmall \rm HHO}(\mbox{lbl}) \nonumber \\
       &=& 350.1 \, (4.8) \times 10^{-8}.
\label{eq:THAD}
\eea
Errors were added in quadrature.

%---------------------------------------------------------------------------%
\section{THE STANDARD MODEL PREDICTION FOR \boldmath $a_{\tau}$ \unboldmath}
%---------------------------------------------------------------------------%
\label{sec:SM}

We can now add up all the contributions discussed in the previous sections
to derive the {\small SM} prediction for $a_{\tau}$~\cite{EP06}:
\bea
    a_{\tau}^{\mysmall \rm SM} &=& 
         a_{\tau}^{\mysmall \rm QED} +
         a_{\tau}^{\mysmall \rm EW}  +
         a_{\tau}^{\mysmall \rm HLO}  +
         a_{\tau}^{\mysmall \rm HHO} \nonumber \\
         &=&117 \, 721 \, (5) \times 10^{-8}.  
\label{eq:nsm}
\eea
Errors were added in quadrature.
%

%--- EXP

The most stringent limit on the anomalous magnetic moment of the
$\tau$ lepton was derived in 2004 by the {\small DELPHI} collaboration from
$e^+e^- \to e^+e^-\tau^+\tau^-$ total cross section measurements at $\sqrt
s$ between 183 and 208 GeV at {\small LEP2}~\cite{delphi}:
\be
                         -0.052 < a_{\tau} < 0.013
\label{eq:exp_delphi1}
\ee
at 95\% confidence level. 
Comparing this result with \eq{nsm}, 
it is clear that the sensitivity of the best existing
measurements is still more than an order of magnitude worse than needed. 
For other limits on $a_{\tau}$ see Ref.~\cite{PDG06,arcadi}.

%---------------------------------------------------------------------------%
\section{CONCLUSIONS}
%---------------------------------------------------------------------------%
\label{sec:CONC}

The $g$$-$$2$ of the
$\tau$ lepton is much more sensitive than the muon one to {\small EW} and
{\small NP} loop effects that give contributions $\sim m_l^2$, making its
measurement an excellent opportunity to unveil (or just constrain) {\small
NP} effects.

Unfortunately, the very short lifetime of the $\tau$ lepton makes it very
difficult to determine its anomalous magnetic moment by measuring its spin
precession in the magnetic field, like in the muon $g$$-$$2$
experiment~\cite{bnl}. Instead, experiments focus on high-precision
measurements of the $\tau$ lepton pair production in various high-energy
processes, comparing the measured cross sections with the {\small
QED} predictions~\cite{delphi,PDG06}, but their sensitivity  is still more 
than an order of magnitude worse than that required to determine $a_{\tau}$. 

Nonetheless, there are many interesting suggestions to measure $a_{\tau}$, 
e.g., from the radiation amplitude zero in radiative $\tau$
decays~\cite{rad_zero} or from other observables. By employing such methods 
at $B$ factories, one can hope 
to benefit from the possibility to collect very high statistics.
A similar method to
study $a_{\tau}$ using radiative $W$ decays and potentially very high data
samples at {\small LHC} was suggested in Ref.~\cite{samu}.  Yet another
method would use the channeling in a bent crystal similarly to the
measurement of magnetic moments of short-living
baryons~\cite{bent_crystal}. 
In the case of the $\tau$ lepton, it
was suggested to use the decay $B^+ \to \tau^+ \nu_{\tau}$, which would
produce polarized $\tau$ leptons~\cite{Samuel_tau} and was recently
observed~\cite{btau}.
We believe that a detailed feasibility study of such experiments, as well 
as further attempts to improve the accuracy of the theoretical prediction 
for $a_{\tau}$, are quite timely.

%---------------------------------------------------------------------------%
\section*{ACKNOWLEDGMENTS}
%---------------------------------------------------------------------------%
We would like to thank the organizers for their excellent coordination of
this workshop, and in particular Alberto Lusiani for his kind invitation.

%---------------------------------------------------------------------------%

\end{document}